\documentclass[12pt]{article}
\usepackage{graphics}
\usepackage{bm}
\usepackage{graphicx}
\usepackage{epstopdf}
\usepackage{amsmath}
\usepackage{amssymb}
\usepackage{amstext}
\usepackage{amscd}
\usepackage{amsfonts}
\usepackage{appendix}
\usepackage{color}
\usepackage{wasysym}
\usepackage{latexsym}
\DeclareMathAlphabet{\EuFrak}{U}{euf}{m}{n}
\DeclareMathAlphabet{\EuScript}{U}{eus}{m}{n}

\newcommand{\dQ}[1]{\frac{\partial {#1}}{\partial Q_i}}

\newcommand{\dT}[1]{\frac{\partial{#1}}{\partial T}}

\newcommand{\corch}[1]{\left[1+(q-1){#1}\right]}
\newcommand{\nd}{\noindent}

\title{{\bf Verlinde's emergent gravity in an $\boldsymbol{n-}$dimensional,  non-additive Tsallis' scenario}}
\author{{D. J. Zamora$^1$ M. C. Rocca$^1$,
A. Plastino$^1$, and G. L. Ferri$^2$} \\
\small{$^1$ La Plata National University and
Argentina's National Research Council}\\
\small{(IFLP-CCT-CONICET)-C. C. 727, 1900 La Plata - Argentina}\\
$^2$ \small{Fac. de C. Exactas-National University La Pampa,} \\
\small{Peru y Uruguay, Santa Rosa, La Pampa, Argentina}}

\date{\today}

\begin{document}

\maketitle

\begin{abstract}
\noindent This paper brings together four distinct but very important physical notions: 1) Entropic force, 2) Entropy-along-a-curve, 3) Tsallis' q-statistics, and 4) Emergent gravitation. 
We  investigate the non additive, classical
(Tsallis') q-statistical mechanics
 of a phase-space curve in $n$ dimensions (3
dimensions, in particular). We focus attention on an entropic force mechanism that
yields a simple realization of it, being able to mimic interesting effects
such as confinement, hard core, and asymptotic freedom, typical of high energy physics. \\
\nd {\bf Keywords:} Phase-space curves; Entropic force; Tsallis' statistical mechanics

\end{abstract}

\newpage

\renewcommand{\theequation}{\arabic{section}.\arabic{equation}}

\setcounter{equation}{0}

\section{Introduction}

\setcounter{equation}{0}

This paper brings together four distinct but very important physical notions: 1) Entropic force, 2) Entropy-along-a-curve (EAC), 3) Tsallis' q-statistics, and 4) Emergent gravitation. \vskip 3mm
  \nd Since the early 90's, Tsallis' q-statistical mechanics has been employed in variegated scientific scenarios endowed with multiple applications \cite{tsallis,web,pre,epjb1,epjb2,epjb3,epjb4,epjb5,epjb6,epjb7,epjb8,epjb9}. It proved useful for astrophysics, specially for self-gravitating systems \cite{PP93,chava,lb,rosa}.
 It
has generated several thousands of both authors and manuscripts  \cite{web}.  Its success reasserts
the notion that a great deal of physics is to be accrued to just statistical considerations.
 An example lies in its application to high
energy physics, where  q-statistics seems to reasonably describe the transverse
momentum distributions of different hadrons \cite{tp11o,tp11,phenix}. In this paper we discuss Tsalllis' scenario for emergent gravity.\vskip 3mm

\nd Emergent gravity   (also called entropic gravity) is the notion that describes gravitation  as an entropic force with macro-scale homogeneity,  but subject to quantum-level disorder. It claims that gravity is not a fundamental interaction. The theory, advanced by Verlinde \cite{verlinde} is based on string theory, black hole physics, and quantum information theory. Here we consider emergent gravity in a classical statistical context.

\vskip 3mm \nd In \cite{curves} we introduced the analysis of the thermal properties of an entropy-along-a-curve (EAC), this is, we studied the classical statistical mechanics of a phase space curve. This unveiled a mechanism
that, via an associated entropic force, provided us with a simple realization of effects
such as confinement, hard core, and asymptotic freedom, along with emergent gravitation. Additionally, we obtained negative
specific heats, a distinctive feature of self-gravitating systems, and negative pressures,
typical of dark energy.  Now, it is well known that gravity strongly depends upon dimensionality. It is very different, for instance in two dimensions than in 3D. Thus, it makes a lot of sense to perform in three     dimensions the analysis of \cite{curves} and we did this in \cite{3d}.

 \nd  The deep discovery of Tsallis’ was to point out that different statistics often uncover new physics, and thousands of researchers followed suit. In this vein, we then investigated in \cite{javier} the q-statistical mechanics of phase space's one-dimensional curves
 associated to the harmonic oscillator hamiltonian. 
We computed Verlinde's entropic force and re-encountered, along with emergent gravitation,   interesting effects analogous
to confinement, hard core and asymptotic freedom of \cite{curves}. \vskip 3mm

\nd  Such effort is extended here to  $n$ dimensional curves, given the above mentioned 
relevance of dimensionality with respect to gravitation. Does the q-scenario affect the properties of emergent gravitation, and in which form? How do $n$ and $q$ interact? To answer these questions is the rationale of the present work.

\section{Preliminaries}

\setcounter{equation}{0}

\nd  There is a strong link between the harmonic oscillator (HO) and the Kepler problem \cite{kepler}. Fung 
worked out the full correspondence between the Kepler problem and the isotropic harmonic
oscillator in Newtonian mechanics by means of a special transformation. He then applied
this to get all the details of the Kepler problem from the simple solution of the isotropic
harmonic oscillator \cite{kepler}. Thus, it is appropriate to introduce the HO in the present effort. We consider a particle attached to a spring connected to the origin, in thermal contact with a heat bath at the inverse temperature $\beta$. We consider an n-dimensional harmonic oscillator-like Hamiltonian
\begin{equation}
\label{eq2.1}
H(P,Q)=P^2+Q^2,
\end{equation}
\begin{equation}
P^2=P_1^2+P_2^2+...+P_n^2; Q^2=Q_1^2+Q_2^2+...+Q_n^2,
\end{equation}
where $P^2$ and $Q^2$ have the dimensions of $H$.  We use Tsallis'
statistical mechanics \cite{tsallis}, in which the probability
distributions are \textit{q-exponentials} \cite{tsallis}
\begin{equation}
e_q(x)=[1+(1-q)x]^\frac{1}{1-q}.
\end{equation}
The partition function in Tsallis statistical is defined as \cite{tsallis}
\begin{equation}
Z(\beta)=\int_{-\infty}^{\infty} {\corch{\beta H(P,Q)}}^{\frac{1}{1-q}} d^nPd^nQ,
\end{equation}
where $1\leq q<2$ is the non-extensibility parameter. Note that, if $q\rightarrow1$, the partition function reduces to the usual one , that means, the Gibbs-Boltzmann's canonical partition function.
Following the procedure of \cite{3d} one arrives at
\begin{equation}
Z(\beta)=\frac{\pi^n}{\Gamma(n)} \int_{0}^{\infty} U^{n-1} {\corch{\beta U}}^{\frac{1}{1-q}} dU,
\label{ZU}
\end{equation}
where we employed the change of variable
\begin{equation}
U=P^2+Q^2.
\end{equation}
Evaluating (\ref{ZU}) we have
\begin{equation}
Z(\beta)=\frac{\pi^n}{\Gamma(n)}\left[\frac{1}{\beta(q-1)}\right]^n B\left[n,\frac{1}{q-1}-n\right],
\end{equation}
where $B[a,b]$ is the beta function. Thus:
\begin{equation}
Z(\beta)=\left[\frac{\pi}{\beta (q-1)}\right]^n \frac{\Gamma\left(\frac{1}{q-1}-n\right)}{\left(\frac{1}{q-1}-1\right)\Gamma\left(\frac{1}{q-1}-1\right)},
\label{eq2.7}
\end{equation}
where we used this property of the Gamma function:
\begin{equation}
\label{gamma}
\Gamma(n+1)=n\Gamma(n).
\end{equation}
Using (\ref{gamma}) n times, we then have
\begin{equation}
Z(\beta)=\left[\frac{\pi}{\beta (q-1)}\right]^n \frac{1}{\left(\frac{2-q}{q-1}\right)\left(\frac{3-2q}{q-1}\right)...\left(\frac{n+1-nq}{q-1}\right)},
\end{equation}
and, finally:
\begin{equation}
Z(\beta)=\left(\frac{\pi}{\beta}\right)^n \frac{1}{\prod_{i=1}^n(i+1-iq)}.
\label{Zinf}
\end{equation}
If $q\rightarrow1$, then
\begin{equation}
Z\rightarrow \left(\frac{\pi}{\beta}\right)^n.
\end{equation}
Equation (\ref{Zinf}) reduces to the expression obtained in \cite{3d} for the  Gibbs-Boltzmann statistical one.
Also note that if $n=1$, then
\begin{equation}
Z(\beta)=\frac{\pi}{\beta}\frac{1}{(2-q)}.
\end{equation}
Equation (\ref{Zinf}) reduces to the 1-dimensional expression obtained in \cite{javier}.
Similarly the mean value of the energy is defined in Tsallis statistical mechanics like:
\begin{equation}
\label{eq2.13}
<U>(\beta)=\frac{1}{Z}\int_{-\infty}^{\infty} H(P,Q){\corch{\beta H(P,Q)}}^{\frac{1}{1-q}} d^nPd^nQ,
\end{equation}
that leads to
\begin{equation}
<U>(\beta)=\frac{\pi^n}{\Gamma(n)Z(\beta)} \int_{0}^{\infty} U^n{\corch{\beta U}}^{\frac{1}{1-q}} dU.
\label{UU}
\end{equation}
(\ref{UU}) results in
\begin{equation}
\label{eq2.15}
<U>(\beta)=\frac{\pi^n}{\Gamma(n)Z(\beta)} \left[\frac{1}{\beta(q-1)}\right]^{n+1} B\left[n+1,\frac{1}{q-1}-n-1\right],
\end{equation}
which is
\begin{equation}
<U>(\beta)=\frac{n\pi^n}{Z(\beta)[\beta(q-1)]^{n+1}}\frac{\Gamma(\frac{1}{q-1}-n-1)}{\Gamma\left(\frac{1}{q-1}\right)}.
\end{equation}
Using  equation (\ref{eq2.7}) we obtain
\begin{equation}
\label{eq2.17}
<U>(\beta)=\frac{n}{\beta(q-1)}\frac{\Gamma(\frac{1}{q-1}-n-1)}{\Gamma\left(\frac{1}{q-1}-n\right)},
\end{equation}
and employing  (\ref{gamma}), equation (\ref{eq2.17}) can be written as
\begin{equation}
<U>(\beta)=\frac{n}{\beta[(n+2)-(n+1)q]},
\label{Uinf}
\end{equation}
with the restriction $1\leq q<\frac{n+2}{n+1}$ in order to guarantee the non-divergence of $<U>$.
When $q\rightarrow1$ we obtain
\begin{equation}
\label{eq2.19}
<U>\rightarrow\frac{n}{\beta},
\end{equation}
and when $n=1$,
\begin{equation}
\label{eq2.20}
<U>=\frac{1}{\beta(3-2q)}.
\end{equation}
The restriction on $q$ becomes $q<3/2$. Equations (\ref{eq2.19}) and (\ref{eq2.20}) are the expressions of $<U>$ obtained in \cite{3d} and \cite{javier}, respectively.\vskip 3mm

\nd Since the entropy $S$ is \cite{div}
\begin{equation}
S(\beta)=ln_{2-q}Z+Z^{q-1}\beta<U>,
\label{s}
\end{equation}
we obtain for the entropy
\begin{multline}
S(\beta)=\left(\frac{\pi}{\beta}\right)^{n(q-1)}\left[\frac{1}
{\prod_{i=1}^n(i+1-iq)}\right]^{q-1}\left[\frac{1}{q-1}+\frac{n}{(n+2)-(n+1)q}\right]\\
-\frac{1}{q-1}.
\end{multline}

\section{Entropy along a path $\Gamma$}

\setcounter{equation}{0}

\nd   We recapitulate here the main ideas advanced in \cite{curves}. We assume that the system is in contact with a reservoir at the fixed inverse temperature $\beta$. We call $\Gamma$ a phase-space path parametrized by $Q_1$ that starts at the origin and ends at some arbitrary point $(P_1(Q_1^0),...,P_n(Q_1^0),Q_1(Q_1^0),...,Q_n(Q_1^0))$. All the calculations are of a microscopic character. Generalizing the exact differentials-integrands (\ref{ZU}) and (\ref{UU}), we introduce

\begin{equation}
Z(\beta,\Gamma)=\frac{\pi^n}{\Gamma(n)} \int_{\Gamma} U^{n-1} {\corch{\beta U}}^{\frac{1}{1-q}} dU,
\end{equation}

\begin{equation}
<U>(\beta,\Gamma)=\frac{\pi^n}{\Gamma(n)Z(\beta)} \int_{\Gamma} U^n{\corch{\beta U}}^{\frac{1}{1-q}} dU.
\end{equation}
Since $P_i(0)=0$ and $Q_i(0)=0$ so is $U(0,0)=0$. The integrands are exact differentials and they only depend on their end-point $Q_1^0$, Accordingly,

\begin{equation}
Z(\beta,Q_1^0)=\frac{\pi^n}{\Gamma(n)} \int_{0}^{Q_1^0} U^{n-1} {\corch{\beta U}}^{\frac{1}{1-q}} dU,
\label{zq}
\end{equation}

\begin{equation}
<U>(\beta,Q_1^0)=\frac{\pi^n}{\Gamma(n)Z(\beta,Q_1^0)} \int_{0}^{Q_1^0} U^n{\corch{\beta U}}^{\frac{1}{1-q}} dU.
\label{uq}
\end{equation}
Integrating by parts $n$ times one finds

\begin{equation}
Z(\beta,Q_1^0)=\frac{\pi^n}{\beta^n}\left\{\frac{1}{\prod_{i=1}^n(i+1-iq)}-\sum_{j=1}^n\frac{(\beta U)^{n-j}}{(n-j)!}\frac{\corch{\beta U}^{\frac{j+1-jq}{1-q}}}{\prod_{k=1}^j(k+1-kq)}\right\},
\label{ZQ}
\end{equation}
where we write $U$ instead of $U(P(Q_1^0),Q(Q_1^0))$ in order to simplify the notation.
Note that when $Q_1^0\rightarrow\infty$ we recover the relation (\ref{Zinf}).
If $n=1$,

\begin{equation}
Z(\beta,Q_1^0)=\frac{\pi}{\beta(2-q)}\{1-\corch{\beta U}^\frac{2-q}{1-q}\},
\end{equation}
and  we arrive at the expression found in \cite{javier}.
The other interesting limit is  $q\rightarrow1$. Here  we obtain
\begin{equation}
Z(\beta,Q_1^0)=\frac{\pi^n}{\beta^n}-e^{-\beta U}\sum_{s=0}^{n-1}\frac{\pi^n}{s!}\frac{U^s}{\beta^{n-s}},
\end{equation}
where we called $s=n-j$ and  we arrive to that equation already obtained in \cite{3d}.
In the same way, integrating by parts $n+1$ times equation (\ref{uq}), we have for the mean energy
\begin{multline}
<U>(\beta,Q_1^0)=\frac{n\pi^n}{\beta^nZ(\beta,Q_1^0)}\left\{\frac{1}{\prod_{i=1}^{n+1}(i+1-iq)\beta}\right.\\
\left.-\sum_{j=1}^{n+1}\frac{\beta^{n-j}U^{n+1-j}}{(n+1-j)!}\frac{\corch{\beta U}^{\frac{j+1-jq}{1-q}}}{\prod_{k=1}^j(k+1-kq)}\right\}.
\label{UQ}
\end{multline}
Again, we recover equation (\ref{Uinf}) when $Q_1^0\rightarrow\infty$.
Appealing to  equation (\ref{ZQ}) we obtain now
\begin{multline}
<U>(\beta,Q_1^0)=n\left\{\frac{1}{\prod_{i=1}^n(i+1-iq)}-\sum_{j=1}^n\frac{(\beta U)^{n-j}}{(n-j)!}\frac{\corch{\beta U}^{\frac{j+1-jq}{1-q}}}{\prod_{k=1}^j(k+1-kq)}\right\}^{-1}\\
\left\{\frac{1}{\prod_{i=1}^{n+1}(i+1-iq)\beta}-\sum_{j=1}^{n+1}\frac{\beta^{n-j}U^{n+1-j}}{(n+1-j)!}\frac{\corch{\beta U}^{\frac{j+1-jq}{1-q}}}{\prod_{k=1}^j(k+1-kq)}\right\}.
\end{multline}
If $q\rightarrow1$ we recover Ref.  \cite{3d}'s result

\begin{equation}
<U>(\beta,Q_1^0)=\frac{\pi^n}{\beta^nZ(\beta,Q_1^0)}\left\{\frac{n}{\beta}-\sum_{s=0}^n\frac{n\beta^{s-1}U^s}{s!}e^{-\beta U}\right\},
\end{equation}
where this time we called $s=n+1-j$. For $n=1$ we find

\begin{multline}
<U>(\beta,Q_1^0)=\frac{1}{\beta\{1-\corch{\beta U}^\frac{2-q}{1-q}\}}\left\{\beta U \corch{\beta U}^\frac{2-q}{1-q}\right.\\
\left.+\frac{\corch{\beta U}^\frac{3-2q}{1-q}}{3-2q}\right\}.
\end{multline}
This is  the solution for equation (3.6) of \cite{javier}, after  integrating by parts.
Finally, the entropy is expressed by equation (\ref{s}), where we have to replace $Z$ and $<U>$ from Eq. (\ref{ZQ}) and (\ref{UQ}), respectively.

\section{Entropic force and total force}

\setcounter{equation}{0}

\nd   According to \cite{verlinde}, the entropic force is given by
\begin{equation}
F_e=\frac{1}{\beta}\dQ{S}.
\end{equation}
In our case that this is
\begin{equation}
F_e=\frac{Z^{q-2}}{\beta}\dQ{Z}[1+(q-1)\beta<U>]+Z^{q-1}\dQ{<U>},
\label{Fe}
\end{equation}
where
\begin{multline}
\label{4.35}
\dQ{Z}=\frac{\pi^n}{\beta^n}2Q_i\sum_{j=1}^n\frac{\beta^{n-j}}{(n-j)!\prod_{k=1}^j(k+1-kq)}\left\{(j+1-jq)\beta U^{n-j}\right.\\
\left.\corch{\beta U}^{j+\frac{q}{1-q}}-(n-j)U^{n-j-1}\corch{\beta U}^{\frac{j+1-jq}{1-q}}\right\},
\end{multline}
and
\begin{multline}
\label{4.36}
\dQ{<U>}=-\frac{<U>}{Z}\dQ{Z}+\frac{n\pi^n}{Z\beta^n}2Q_i\sum_{j=1}^{n+1}\frac{\beta^{n-j}}{(n+1-j)!\prod_{k=1}^j(k+1-kq)}\\
\left\{(j+1-jq)\beta U^{n+1-j}\corch{\beta U}^{j+\frac{q}{1-q}}\right.\\
\left.-(n+1-j)U^{n-j}\corch{\beta U}^\frac{j+1-jq}{1-q}\right\}.
\end{multline}
Using (\ref{4.35}), (\ref{4.36}),(\ref{Fe})
we obtain the final expression for the entropic force given by
(\ref{a1}) of the Appendix.

\nd   If $n=1$ equations (\ref{4.35}) and (\ref{4.36}) reduce, respectively, to
\begin{equation}
\dQ{Z}=2\pi Q\corch{\beta U}^\frac{1}{1-q},
\end{equation}

\begin{equation}
\dQ{<U>}=-\frac{<U>}{Z}\dQ{Z}+\frac{2QU\pi}{Z}\corch{\beta U}^\frac{1}{1-q}.
\end{equation}
This  is, they reduce to the expressions obtained in \cite{javier}.
They also agree with the equations obtained in \cite{3d} when $q\rightarrow1$
\begin{equation}
\dQ{Z}=2\pi^nQe^{-\beta U}\left\{\sum_{k=0}^{n+1}\frac{U^k}{k!\beta^{n-k-1}}-\sum_{k=0}^{n-1}\frac{U^{k-1}}{(k-1)!\beta^{n-k}}\right\},
\label{4.39}
\end{equation}
where we called $k=n-j$.
\begin{equation}
\dQ{<U>}=-\frac{<U>}{Z}\dQ{Z}+\frac{2Q\pi^n}{Z}e^{-\beta U}\left[\sum_{k=0}^{n}\frac{nU^k}{k!\beta^{n-k}}-\sum_{k=0}^n\frac{nU^{k-1}}{(k-1)!\beta^{n-k+1}}\right],
\label{4.40}
\end{equation}
in which we called $k=n+1-j$. Since equations (\ref{4.35}) and (\ref{4.36}) reduces to (\ref{4.39}) and (\ref{4.40}) and when $q\rightarrow1$ equation (\ref{Fe}) reduces to

\begin{equation}
F_e=\frac{1}{\beta Z}\dQ{Z}+\dQ{<U>},
\end{equation}
and then, the limit of $F_e$ agrees with that of \cite{3d}. The complete form for $F_e$ can be found in 
the Appendix, see Eq. (\ref{a1}).\vskip 3mm

\begin{figure}[h!]
\begin{center}
\includegraphics[width=14.6cm]
{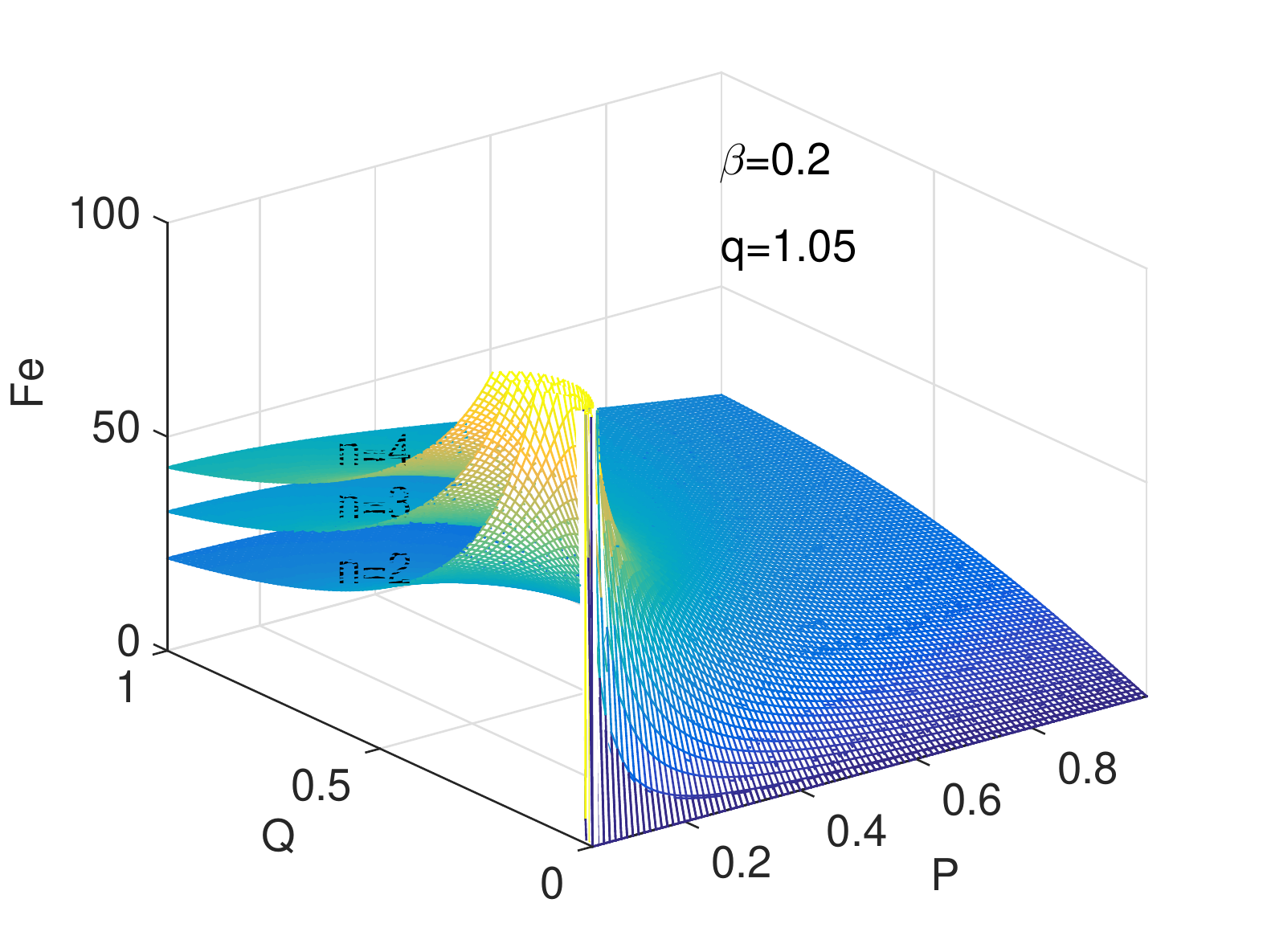} \caption{Behaviour of the entropic force
with $n$. We note that the entropic force grows with the
dimensionality.} \label{figura1}
\end{center}
\end{figure}

\nd Of course the particle also feels the influence of the negative gradient of the HO potential:
\begin{equation}
F_{HO}=-\frac{1}{2}\dQ{<U>},
\end{equation}
whose expression is given by (\ref{a2}) of the Appendix.
Note that the force vanishes at the origin, as indicated by the graph. The force of the harmonic oscillator (i) changes sign  in going from one side of the origin to  the other and (ii) depends on whether one is compressing or elongating the associated coil spring. Remember also that we are dealing with the forces' statistical averages.
\vskip 2mm
\nd Now, we are in the presence of a total force
\begin{equation}
F_T=F_e+F_{HO}.
\end{equation}

\begin{figure}[h!]
\begin{center}
\includegraphics[width=14.6cm]
{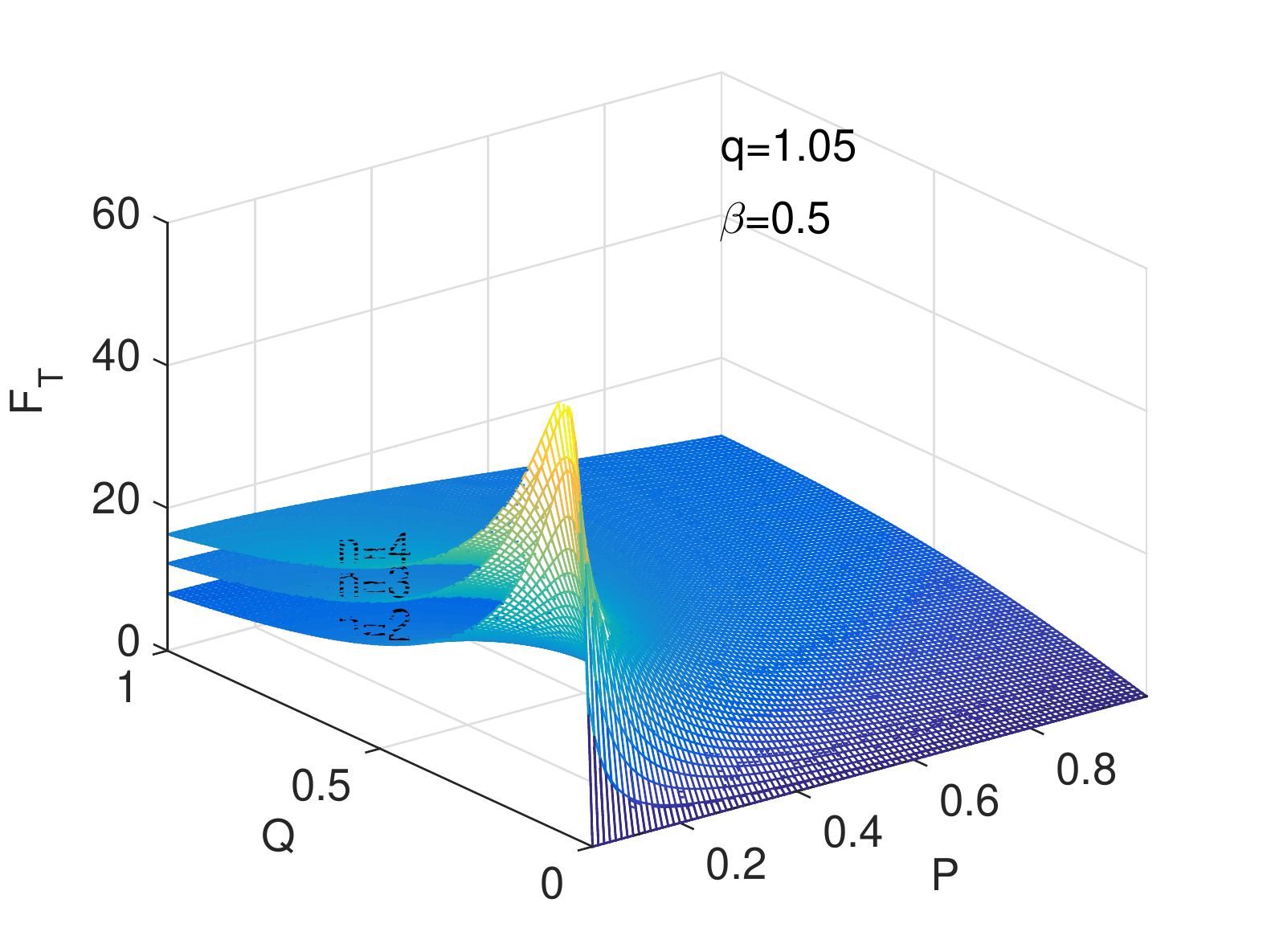} \caption{Behaviour of the total force
with $n$. We note that the total force grows with the
dimensionality.} \label{figura2}
\end{center}
\end{figure}

\nd   One has
\begin{equation}
F_T=\frac{1}{\beta}Z^{q-2}\dQ{Z}[1+(q-1)\beta <U>]+\dQ{<U>}\left(Z^{q-1}-\frac{1}{2}\right),
\end{equation}
and from this one is led to the form for $F_T$ of Eq. (\ref{a3}) in the Appendix.
\section{Specific heat}

\setcounter{equation}{0}

\nd   The specific heat can be calculated as
\begin{multline}
C=\dT{<U>}=-\frac{<U>}{Z}\dT{Z}+\frac{k_Bn\pi^n}{Z}\left\{\frac{n+1}{\beta^n\prod_{i=1}^n(i+1-iq)}\right.\\
\left.-\sum_{j=1}^{n+1}\frac{U^{n+1-j}}{(n+1-j)!\prod_{k=1}^j(k+1-kq)}\left[\frac{j}{\beta^{j-1}}\corch{\beta U}^\frac{j+1-jq}{1-q}\right.\right.\\
\left.\left.+\frac{(j+1-jq)U}{\beta^{j-2}}\corch{\beta U}^{j+\frac{q}{1-q}}\right]\right\},
\end{multline}
where
\begin{multline}
\dT{Z}=\frac{nk_B\pi^n}{\beta^{n-1}\prod_{i=1}^n(i+1-iq)}-k_B\pi^n\sum_{j=1}^n\frac{U^{n-j}}{(n-j!)\prod_{k=1}^j(k+1-kq)}\\
\left\{\frac{j}{\beta^{j-1}}\corch{\beta U}^\frac{j+1-jq}{1-q}+\frac{(j+1-jq)U}{\beta^{j-2}}\corch{\beta U}^{j+\frac{q}{1-q}}\right\}.
\end{multline}
Thus on obtains for $C$ Eq. (\ref{a5}) of the Appendix.  
When $q\rightarrow1$
\begin{multline}
\dT{Z}=k\pi^n\left[n(kT)^{n-1}-e^{-\beta U}\sum_{s=0}^{n-1}\frac{U^{s+1}}{s!}(kT)^{n-s-2}\right.\\
\left.-e^{-\beta U}\sum_{s=0}^{n-1}\frac{(n-s)U^s}{s!}(kT)^{n-s-1}]\right],
\end{multline}
 which agrees  with the result given in \cite{3d}. In the expression above we called $s=n-j$.
 When $n=1, \dT{Z}$ becomes
\begin{equation}
\dT{Z}=\frac{k\pi}{(2-q)}\{1-\corch{\beta U}^\frac{2-q}{1-q}+(2-q)\beta U\corch{\beta U}^\frac{1}{1-q}\},
\end{equation}
as in \cite{javier}. The limit of $C$ when $q\rightarrow1$ also coincide with \cite{3d}:

\begin{multline}
C=-\frac{<U>}{Z}\dT{Z}+\frac{k\pi^n}{Z}\left[\frac{n(n+1)}{\beta^n}-\sum_{s=0}^n\frac{nU^{s+1}}{s!\beta^{n-s-1}}-e^{-\beta U}\sum_{s=0}^n\frac{n(n-s+1)U^s}{s!\beta^{n-s}}\right]
\end{multline}

\section{Conclusions}

\setcounter{equation}{0}

\nd We have here brought together four disparate but very important physical notions: 1) Entropic force, 2) Entropy-along-a-curve, 3) Tsallis q-statistics, and 4) Emergent gravitation in discussing the q-statistics of phase space curves.\vskip 3mm 

\nd   The expressions for all the quantities here discussed are generalizations of those found in \cite{3d} and
\cite{javier} for the limits $q\rightarrow1$ and $n=1$, respectively. This reconfirms their correctness. The behaviour is the same in all dimensions, which includes hard-core, confinement and asymptotic freedom. This should motivate efforts directed to find these properties in the Lab, in more general settings than those of high energy physics.\vskip 3mm

\nd Interestingly enough, although dimensionality is very important in gravity's workings, in what respects to phase space curves, we proved that, {\it qualitatively}, the statistical properties of them, e.g.,
the entropic forces, are {\it intrinsic} to the curve, no matter in
what space it is embedded. {\it Quantitatively, our graphs show that the entropic force grows with the dimensionality $n$}. This can be understood if we accept that disorder grows with $n$, as the number of concomitant spatial ''arrangements'' of any kind augments with $n$, and so that the entropy variationa along a curve.

\newpage

\renewcommand{\thesection}{\Alph{section}}

\renewcommand{\theequation}{\Alph{section}.\arabic{equation}}

\setcounter{section}{1}

\section*{Appendix A}

\setcounter{equation}{0}
\nd We give here the explicit form of some rather complicated equations dealt with in the text. For the entropic force we have
\begin{multline}
\label{a1}
F_e=\frac{2Q_i}{\beta}\frac{\pi^{n(q-1)}}{\beta^{n(q-1)}}\left\{\frac{1}{\prod_{i=1}^n(i+1-iq)}-\sum_{j=1}^n\frac{(\beta U)^{n-j}}{(n-j)!}\frac{\corch{\beta U}^{\frac{j+1-jq}{1-q}}}{\prod_{k=1}^j(k+1-kq)}\right\}^{q-2}\\
\sum_{j=1}^n\frac{\beta^{n-j}}{(n-j)!\prod_{k=1}^j(k+1-kq)}\left\{(j+1-jq)\beta U^{n-j}\corch{\beta U}^{j+\frac{q}{1-q}}\right.\\
\left.-(n-j)U^{n-j-1}\corch{\beta U}^{\frac{j+1-jq}{1-q}}\right\}\left[1+(q-1)\beta n\left\{\frac{1}{\prod_{i=1}^n(i+1-iq)}\right.\right.\\
\left.\left.-\sum_{j=1}^n\frac{(\beta U)^{n-j}}{(n-j)!}\frac{\corch{\beta U}^{\frac{j+1-jq}{1-q}}}{\prod_{k=1}^j(k+1-kq)}\right\}^{-1}\left\{\frac{1}{\prod_{i=1}^{n+1}(i+1-iq)\beta}\right.\right.\\
\left.\left.-\sum_{j=1}^{n+1}\frac{\beta^{n-j}U^{n+1-j}}{(n+1-j)!}\frac{\corch{\beta U}^{\frac{j+1-jq}{1-q}}}{\prod_{k=1}^j(k+1-kq)}\right\}\right]+\frac{\pi^{n(q-1)}}{\beta^{n(q-1)}}\left\{\frac{1}{\prod_{i=1}^n(i+1-iq)}\right.\\
\left.-\sum_{j=1}^n\frac{(\beta U)^{n-j}}{(n-j)!}\frac{\corch{\beta U}^{\frac{j+1-jq}{1-q}}}{\prod_{k=1}^j(k+1-kq)}\right\}^{q-1}\left\{-2nQ_i\left\{\frac{1}{\prod_{i=1}^n(i+1-iq)}\right.\right.\\
\left.\left.-\sum_{j=1}^n\frac{(\beta U)^{n-j}}{(n-j)!}\frac{\corch{\beta U}^{\frac{j+1-jq}{1-q}}}{\prod_{k=1}^j(k+1-kq)}\right\}^{-1}\left\{\frac{1}{\prod_{i=1}^{n+1}(i+1-iq)\beta}\right.\right.\\
\left.\left.-\sum_{j=1}^{n+1}\frac{\beta^{n-j}U^{n+1-j}}{(n+1-j)!}\frac{\corch{\beta U}^{\frac{j+1-jq}{1-q}}}{\prod_{k=1}^j(k+1-kq)}\right\}\left\{\frac{1}{\prod_{i=1}^n(i+1-iq)}\right.\right.\\
\left.\left.-\sum_{j=1}^n\frac{(\beta U)^{n-j}}{(n-j)!}\frac{\corch{\beta U}^{\frac{j+1-jq}{1-q}}}{\prod_{k=1}^j(k+1-kq)}\right\}^{-1}\sum_{j=1}^n\frac{\beta^{n-j}}{(n-j)!\prod_{k=1}^j(k+1-kq)}\right.\\
\left.\left\{(j+1-jq)\beta U^{n-j}\corch{\beta U}^{j+\frac{q}{1-q}}-(n-j)U^{n-j-1}\corch{\beta U}^{\frac{j+1-jq}{1-q}}\right\}\right.\\
\left.+2nQ_i\left\{\frac{1}{\prod_{i=1}^n(i+1-iq)}-\sum_{j=1}^n\frac{(\beta U)^{n-j}}{(n-j)!}\frac{\corch{\beta U}^{\frac{j+1-jq}{1-q}}}{\prod_{k=1}^j(k+1-kq)}\right\}^{-1}\right.\\
\left.\sum_{j=1}^{n+1}\frac{\beta^{n-j}}{(n+1-j)!\prod_{k=1}^j(k+1-kq)}\left\{(j+1-jq)\beta U^{n+1-j}\corch{\beta U}^{j+\frac{q}{1-q}}\right.\right.\\
\left.\left.-(n+1-j)U^{n-j}\corch{\beta U}^\frac{j+1-jq}{1-q}\right\}\right\}.
\end{multline}
For the HO force one has
\begin{multline}
\label{a2}
F_{HO}=nQ_i\left\{\frac{1}{\prod_{i=1}^n(i+1-iq)}-\sum_{j=1}^n\frac{(\beta U)^{n-j}}{(n-j)!}\frac{\corch{\beta U}^{\frac{j+1-jq}{1-q}}}{\prod_{k=1}^j(k+1-kq)}\right\}^{-2}\\
\left\{\frac{1}{\prod_{i=1}^{n+1}(i+1-iq)\beta}-\sum_{j=1}^{n+1}\frac{\beta^{n-j}U^{n+1-j}}{(n+1-j)!}\frac{\corch{\beta U}^{\frac{j+1-jq}{1-q}}}{\prod_{k=1}^j(k+1-kq)}\right\}\\
\sum_{j=1}^n\frac{\beta^{n-j}}{(n-j)!\prod_{k=1}^j(k+1-kq)}\left\{(j+1-jq)\beta U^{n-j}\right.\\
\left.\corch{\beta U}^{j+\frac{q}{1-q}}-(n-j)U^{n-j-1}\corch{\beta U}^{\frac{j+1-jq}{1-q}}\right\}\\
-nQ_i\left\{\frac{1}{\prod_{i=1}^n(i+1-iq)}-\sum_{j=1}^n\frac{(\beta U)^{n-j}}{(n-j)!}\frac{\corch{\beta U}^{\frac{j+1-jq}{1-q}}}{\prod_{k=1}^j(k+1-kq)}\right\}^{-1}\\
\sum_{j=1}^{n+1}\frac{\beta^{n-j}}{(n+1-j)!\prod_{k=1}^j(k+1-kq)}\left\{(j+1-jq)\beta U^{n+1-j}\corch{\beta U}^{j+\frac{q}{1-q}}\right.\\
\left.-(n+1-j)U^{n-j}\corch{\beta U}^\frac{j+1-jq}{1-q}\right\}.
\end{multline}
The total force is
\begin{multline}
\label{a3}
F_T=\frac{2Q_i}{\beta}\frac{\pi^{n(q-1)}}{\beta^{n(q-1)}}\left\{\frac{1}{\prod_{i=1}^n(i+1-iq)}-\sum_{j=1}^n\frac{(\beta U)^{n-j}}{(n-j)!}\frac{\corch{\beta U}^{\frac{j+1-jq}{1-q}}}{\prod_{k=1}^j(k+1-kq)}\right\}^{q-2}\\
\sum_{j=1}^n\frac{\beta^{n-j}}{(n-j)!\prod_{k=1}^j(k+1-kq)}\left\{(j+1-jq)\beta U^{n-j}\corch{\beta U}^{j+\frac{q}{1-q}}\right.\\
\left.-(n-j)U^{n-j-1}\corch{\beta U}^{\frac{j+1-jq}{1-q}}\right\}\left[1+(q-1)\beta n\left\{\frac{1}{\prod_{i=1}^n(i+1-iq)}\right.\right.\\
\left.\left.-\sum_{j=1}^n\frac{(\beta U)^{n-j}}{(n-j)!}\frac{\corch{\beta U}^{\frac{j+1-jq}{1-q}}}{\prod_{k=1}^j(k+1-kq)}\right\}^{-1}\left\{\frac{1}{\prod_{i=1}^{n+1}(i+1-iq)\beta}-\sum_{j=1}^{n+1}\frac{\beta^{n-j}U^{n+1-j}}{(n+1-j)!}\right.\right.\\
\left.\left.\frac{\corch{\beta U}^{\frac{j+1-jq}{1-q}}}{\prod_{k=1}^j(k+1-kq)}\right\}\right]+\frac{\pi^{n(q-1)}}{\beta^{n(q-1)}}\left\{\frac{1}{\prod_{i=1}^n(i+1-iq)}-\sum_{j=1}^n\frac{(\beta U)^{n-j}}{(n-j)!}\frac{\corch{\beta U}^{\frac{j+1-jq}{1-q}}}{\prod_{k=1}^j(k+1-kq)}\right\}^{q-1}\\
\left\{-2nQ_i\left\{\frac{1}{\prod_{i=1}^n(i+1-iq)}-\sum_{j=1}^n\frac{(\beta U)^{n-j}}{(n-j)!}\frac{\corch{\beta U}^{\frac{j+1-jq}{1-q}}}{\prod_{k=1}^j(k+1-kq)}\right\}^{-1}\left\{\frac{1}{\prod_{i=1}^{n+1}(i+1-iq)\beta}\right.\right.\\
\left.\left.-\sum_{j=1}^{n+1}\frac{\beta^{n-j}U^{n+1-j}}{(n+1-j)!}\frac{\corch{\beta U}^{\frac{j+1-jq}{1-q}}}{\prod_{k=1}^j(k+1-kq)}\right\}\left\{\frac{1}{\prod_{i=1}^n(i+1-iq)}-\sum_{j=1}^n\frac{(\beta U)^{n-j}}{(n-j)!}\right.\right.\\
\left.\left.\frac{\corch{\beta U}^{\frac{j+1-jq}{1-q}}}{\prod_{k=1}^j(k+1-kq)}\right\}^{-1}\sum_{j=1}^n\frac{\beta^{n-j}}{(n-j)!\prod_{k=1}^j(k+1-kq)}\left\{(j+1-jq)\beta U^{n-j}\corch{\beta U}^{j+\frac{q}{1-q}}\right.\right.\\
\left.\left.-(n-j)U^{n-j-1}\corch{\beta U}^{\frac{j+1-jq}{1-q}}\right\}+2nQ_i\left\{\frac{1}{\prod_{i=1}^n(i+1-iq)}-\sum_{j=1}^n\frac{(\beta U)^{n-j}}{(n-j)!}\right.\right.\\
\left.\left.\frac{\corch{\beta U}^{\frac{j+1-jq}{1-q}}}{\prod_{k=1}^j(k+1-kq)}\right\}^{-1}\sum_{j=1}^{n+1}\frac{\beta^{n-j}}{(n+1-j)!\prod_{k=1}^j(k+1-kq)}\left\{(j+1-jq)\beta U^{n+1-j}\right.\right.\\
\left.\left.\corch{\beta U}^{j+\frac{q}{1-q}}-(n+1-j)U^{n-j}\corch{\beta U}^\frac{j+1-jq}{1-q}\right\}\right\}+nQ_i\left\{\frac{1}{\prod_{i=1}^n(i+1-iq)}\right.\\
\end{multline}
\[\left.-\sum_{j=1}^n\frac{(\beta U)^{n-j}}{(n-j)!}\frac{\corch{\beta U}^{\frac{j+1-jq}{1-q}}}{\prod_{k=1}^j(k+1-kq)}\right\}^{-2}\left\{\frac{1}{\prod_{i=1}^{n+1}(i+1-iq)\beta}-\sum_{j=1}^{n+1}\frac{\beta^{n-j}U^{n+1-j}}{(n+1-j)!}\right.\]
\[\left.\frac{\corch{\beta U}^{\frac{j+1-jq}{1-q}}}{\prod_{k=1}^j(k+1-kq)}\right\}\sum_{j=1}^n\frac{\beta^{n-j}}{(n-j)!\prod_{k=1}^j(k+1-kq)}\left\{(j+1-jq)\beta U^{n-j}\right.\]
\[\left.\corch{\beta U}^{j+\frac{q}{1-q}}-(n-j)U^{n-j-1}\corch{\beta U}^{\frac{j+1-jq}{1-q}}\right\}-nQ_i\left\{\frac{1}{\prod_{i=1}^n(i+1-iq)}\right.\]
\[\left.-\sum_{j=1}^n\frac{(\beta U)^{n-j}}{(n-j)!}\frac{\corch{\beta U}^{\frac{j+1-jq}{1-q}}}{\prod_{k=1}^j(k+1-kq)}\right\}^{-1}\sum_{j=1}^{n+1}\frac{\beta^{n-j}}{(n+1-j)!\prod_{k=1}^j(k+1-kq)}\]
\[\left\{(j+1-jq)\beta U^{n+1-j}\corch{\beta U}^{j+\frac{q}{1-q}}-(n+1-j)U^{n-j}\corch{\beta U}^\frac{j+1-jq}{1-q}\right\}\]. 
Finally, for the specific heat we have
\begin{multline}
\label{a5}
C=\frac{-n\beta^n}{\pi^n}\left\{\frac{1}{\prod_{i=1}^n(i+1-iq)}-\sum_{j=1}^n\frac{(\beta U)^{n-j}}{(n-j)!}\frac{\corch{\beta U}^{\frac{j+1-jq}{1-q}}}{\prod_{k=1}^j(k+1-kq)}\right\}^{-2}\\
\left\{\frac{1}{\prod_{i=1}^{n+1}(i+1-iq)\beta}-\sum_{j=1}^{n+1}\frac{\beta^{n-j}U^{n+1-j}}{(n+1-j)!}\frac{\corch{\beta U}^{\frac{j+1-jq}{1-q}}}{\prod_{k=1}^j(k+1-kq)}\right\}\\
\left\{\frac{nk_B\pi^n}{\beta^{n-1}\prod_{i=1}^n(i+1-iq)}-k_B\pi^n\sum_{j=1}^n\frac{U^{n-j}}{(n-j!)\prod_{k=1}^j(k+1-kq)}\right.\\
\left.\left\{\frac{j}{\beta^{j-1}}\corch{\beta U}^\frac{j+1-jq}{1-q}+\frac{(j+1-jq)U}{\beta^{j-2}}\corch{\beta U}^{j+\frac{q}{1-q}}\right\}\right\}\\
+k_Bn\beta^n\left\{\frac{1}{\prod_{i=1}^n(i+1-iq)}-\sum_{j=1}^n\frac{(\beta U)^{n-j}}{(n-j)!}\frac{\corch{\beta U}^{\frac{j+1-jq}{1-q}}}{\prod_{k=1}^j(k+1-kq)}\right\}^{-1}\\
\left\{\frac{n+1}{\beta^n\prod_{i=1}^n(i+1-iq)}-\sum_{j=1}^{n+1}\frac{U^{n+1-j}}{(n+1-j)!\prod_{k=1}^j(k+1-kq)}\right.\\
\left.\left[\frac{j}{\beta^{j-1}}\corch{\beta U}^\frac{j+1-jq}{1-q}+\frac{(j+1-jq)U}{\beta^{j-2}}\corch{\beta U}^{j+\frac{q}{1-q}}\right]\right\}.
\end{multline}
\end{document}